\documentclass[aps,twocolumn,showpacs,floatfix]{revtex4}
\usepackage[T1]{fontenc}
\usepackage{graphicx}
\usepackage{amstext}
\usepackage{amsxtra}

\begin{document}
\title{Observation and measurement of an extra phase shift created by optically detuned light storage in metastable helium}

\author{M.-A. Maynard$^1$}
\author{T. Labidi$^1$}
\author{M. Mukhtar$^1$}
\author{S. Kumar$^2$}
\author{R. Ghosh$^{2,3}$}
\author{F. Bretenaker$^1$}
\author{F. Goldfarb$^1$}
\email{fabienne.goldfarb@u-psud.fr}
\affiliation{$^1$Laboratoire Aim\'e Cotton, CNRS - Universit\'e Paris Sud 11 - ENS Cachan, 91405 Orsay Cedex, France\\
$^2$School of Physical Sciences, Jawaharlal Nehru University, New Delhi 110067, India\\
$^3$School of Natural Sciences, Shiv Nadar University, Gautam Budh Nagar, UP 203207, India}


\date{\today}
\begin{abstract}
Electromagnetically induced transparency (EIT) in metastable helium at room temperature is experimentally shown to exhibit light storage capabilities for intermediate values of the detuning between the coupling and probe beams and the center of the atomic Doppler profiles. An additional phase shift is shown to be imposed to the retrieved pulse of light when the EIT protocol is performed at non-zero optical detunings. The value of this phase shift is measured for different optical detunings between 0 and 2\,GHz, and its origin is discussed.
\end{abstract}

\maketitle

Since the discovery of coherent population trapping (CPT) \cite{Alzetta1976,Arimondo1976} and electromagnetically induced transparency (EIT) 15 years later \cite{Boller1991}, different applications of these phenomena were found, ranging from atom cooling \cite{Aspect1988} to the decrease of group velocities  down to a few meters per second \cite{Hau1999,Kash1999}. This led to the well-known stored light experiments that were performed using EIT in various systems such as cold atoms, gas cells, or doped crystals \cite{Liu2001,Phillips2001,Hedges2010}.

The EIT-based storage protocol relies on the long-lived Raman coherence between the two fundamental states of a $\Lambda$ system, where two ground levels are optically coupled to the same excited level. When a strong coupling beam is applied on one of the two transitions, a narrow transparency window limited by the Raman coherence decay rate is opened along the other leg of the $\Lambda$ system. Because of the slow-light effect associated with such a dramatic change of the absorption properties of the medium, a weak probe pulse that excites the second transition is compressed when it propagates inside the medium. When it is fully inside this medium, the coupling beam can be suddenly switched off and the signal is then mapped onto the Raman coherences that were excited by the two photon process. Finally, the signal pulse can be simply retrieved by switching on the coupling beam again.

Atoms at room temperature in a gas cell are particularly attractive for light storage because of the simplicity of their implementation. However, the significant Doppler broadening has to be considered and might be expected to place a strong limitation. Its effect can nevertheless be minimized using co-propagating coupling and probe beams, and easy-of-use simple gas cells have thus turned out to be attractive for slow or even stopped light experiments \cite{Novikova2012}. The atoms preferably used for such experiments are alkali atoms, mainly rubidium and sometimes sodium or caesium. Experimental achievements have shown that squeezing can be preserved after slowing down or storage by EIT at optical resonance in an alkali cell at room temperature \cite{Hetet2008b,Appel2008}, and the same phenomenon was used to store and retrieve Laguerre-Gaussian modes \cite{Pugatch2007} and 2D images \cite{Vudyasetu2008,Shuker2008}, or to entangle 2D images \cite{Boyer2008}. Quite good results were also obtained with caesium atoms in the Raman regime \cite{Reim2011},\cite{Reim2012} -- with an optical detuning more than 10 times larger than the Doppler width --  and some previous results were obtained with other alkali atoms and smaller detunings of about the Doppler width \cite{Eisaman2005a,Appel2008,Vudyasetu2008}.

In the present paper we present experimental results obtained not with alkali but with metastable helium atoms at room temperature. We could explore EIT-based light storage in the intermediate detuning regime, i.\,e., with coupling and probe beams slightly optically detuned from the center of Doppler profile while maintaining the two photon resonance condition. The detunings that we explore range from 0 to twice the Doppler linewidth, which is possible thanks to the simple level structure of metastable $^4$He, which is free of hyperfine splitting. The large separation between the D0 and D1 lines opens the unique possibility to operate in such intermediate detuning configurations and achieve a systematic study of the effect of the optical detuning the storage efficiency. In particular, this allows us to isolate a phase shift that is accumulated by light during the storage and retrieval process when operating out of optical resonance. The evolution of this phase shift with the optical detuning is investigated for the first time, to the best of our knowledge. Possible applications for experiments performed with large detunings and for quantum information processing are discussed. 

\begin{figure}
\includegraphics[width=0.49\textwidth]{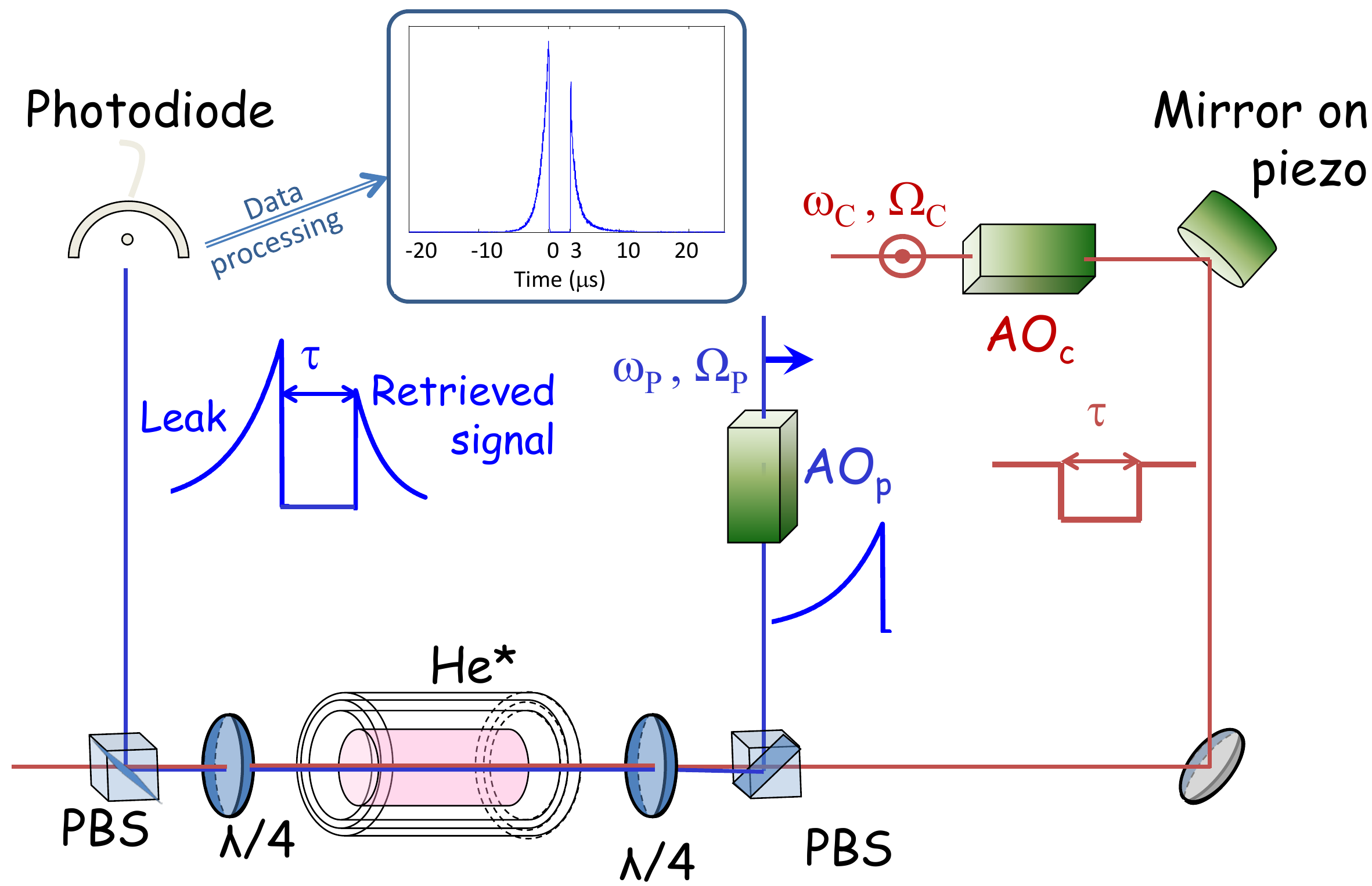} 
\caption{\label{fig1} Experimental setup for EIT storage in metastable $^4$He. The perpendicular linearly polarized coupling and probe beams of Rabi frequencies $\Omega_c$ and $\Omega_p$, respectively, are controlled in frequency and amplitude by acousto-optic modulators (AOs). The helium cell is protected from stray magnetic fields by a $\mu$-metal shielding. A $\lambda/4$ plate and a polarizing beam splitter (PBS) separate the coupling from the probe beam at the output of the cell. Detection is performed by a photodiode. One of the mirrors reflecting only the coupling beam is mounted on a piezoelectric transducer, in order to modulate the phase of the homodyne detection. The recorded signal shows a leak followed by a retrieved pulse after a storage time $\tau$. Inset: Recorded leak and retrieved signals after data processing. When the coupling is switched on again after a 3\,$\mu$s storage time, a reversed exponential retrieved signal is transmitted. The storage efficiency is the ratio of the area under the retrieved signal and the area under a transmitted probe when the cell is switched off.}
\end{figure}

The experiment is based on the $2^3\mathrm{S}_1 \rightarrow 2^3\mathrm{P}_1$ (D1) transition of helium, that permits one to isolate a pure $\Lambda$ system involving only electronic spins. It has previously been shown that it can exhibit very narrow EIT resonances with a kilohertz linewidth, leading to dramatic changes in the group velocity of the probe beam \cite{Goldfarb2008,Goldfarb2009}. Such results are extremely promising when it comes to storage experiments. Figure \ref{fig1} gives the schematic of the experimental set-up. The cell, filled with 1\,Torr of $^4$He, is 6\,cm long and is isolated from magnetic field inhomogeneities thanks to a three-layer $\mu$-metal shield. In these conditions, the Doppler broadened transition half width at half maximum is about 0.9\,GHz, but the optical pumping, which is redistributed by velocity changing collisions, was shown to be effective over approximately half of the Doppler profile \cite{Goldfarb2009}. Helium atoms are excited to the metastable state by an RF discharge at 27 MHz. The linear transmission of a small probe beam is measured to lie between 0.1\% and 0.15\%, depending on the RF discharge conditions: this measurement is performed in the presence of the pumping beam, but with a probe detuned by 1\,MHz to be out of the EIT window.

The coupling and probe beams are derived from the same laser diode at 1.083 $\mu$m. They are controlled in frequency and amplitude by two acousto-optic modulators (AOs), and recombined with a polarizing beam-splitter (PBS). The coupling power can be varied up to 20\,mW. The probe power is always less than 150\,$\mu$W in the experiments described below. A quarter-wave plate ($\lambda/4$) located at the entrance of the cell changes the linear and perpendicular coupling and probe beam polarizations into circular and orthogonal ($\sigma\perp\sigma$) polarizations. After the cell, polarization optics allow detection of only the probe beam. The beam diameters are about 3 mm inside the cell: measurements of the evolution of the width of the EIT transparency window with the coupling power in these conditions give a Raman coherence decay rate $\Gamma_R/2\pi\approx14$\,kHz. As the Raman coherence decay rate was shown to be a few kHz with larger beams, it now gets limited by the transit time of the atoms through the laser beam \cite{Gilles2001}. For a coupling power of about 17\,mW at the cell entrance, the EIT resonance width at resonance was more than 500\,kHz and the maximum transmission of a bit less than 30\%.

The storage protocol that we use is as follows. The coupling beam is first switched on.  The probe beam is then progressively turned on, following an exponential profile followed by an abrupt decrease. Once this probe pulse has entered the helium cell, the coupling beam is suddenly switched off. After a storage time $\tau$ that can be varied, the coupling beam is suddenly switched on and the retrieved pulse is released by the atoms.

It is important to notice that the coupling beam is not totally eliminated by the polarization optics. To determine the retrieved pulse intensity, the remaining coupling intensity cannot be simply subtracted from the total intensity because of the existence of an interference term between probe and coupling. We took advantage of this effect to perform a homodyne-like detection, using the leakage of the coupling beam as a local oscillator. The phase of this local oscillator is slowly scanned using a mirror mounted on a piezo-actuator which reflects the coupling beam only (see fig. \ref{fig1}). The detected signal is accumulated for many different values of the local oscillator phase, and the upper and lower parts of the recorded envelope correspond to relative phases between the coupling and probe beams given by $\Delta\varphi=0 \ [2\pi]$ and $\Delta\varphi=\pi \ [2\pi]$. Knowing the coupling intensity $I_C$ and using the two beam interference formula $I_C+I_P+2\sqrt{I_C I_P}\cos(\Delta\varphi)$, one can deduce the probe intensity $I_P$.  However, one has to take into account the imperfect contrast of the two beams due to a possible small angle between them and to their non-planar wavefronts. In the formula that we use, the factor 2 in the interference term is replaced by a factor $\alpha$: $I_C+I_P+\alpha\sqrt{I_C I_P}\cos(\Delta\varphi)$, $\alpha$ being measured for each set of data and found to be larger than 1.65 for the results reported here. If one considers only the effect of misalignments, such an $\alpha$ factor corresponds to angles smaller than 0.01$^{\circ}$.

The inset in fig.\,\ref{fig1} shows a typical recorded signal after data processing. As explained above, the probe pulse leading edge is taken as exponential, in order to better fit the Lorentzian frequency spectrum of the EIT resonance. The probe power at its maximum is about 130\,$\mu$W. The first detected peak is the leak that is transmitted through the cell, due to its finite absorption. The storage time is 3\,$\mu$s in this example. Once the coupling beam has been switched on again, the retrieved signal is clearly visible. The storage efficiency is measured as the ratio of the area under the retrieved signal to the area under a transmitted probe recorded without metastable helium atoms, i.\,e. with the RF discharge switched off. 

\begin{figure}
\center
\includegraphics[width=0.45\textwidth]{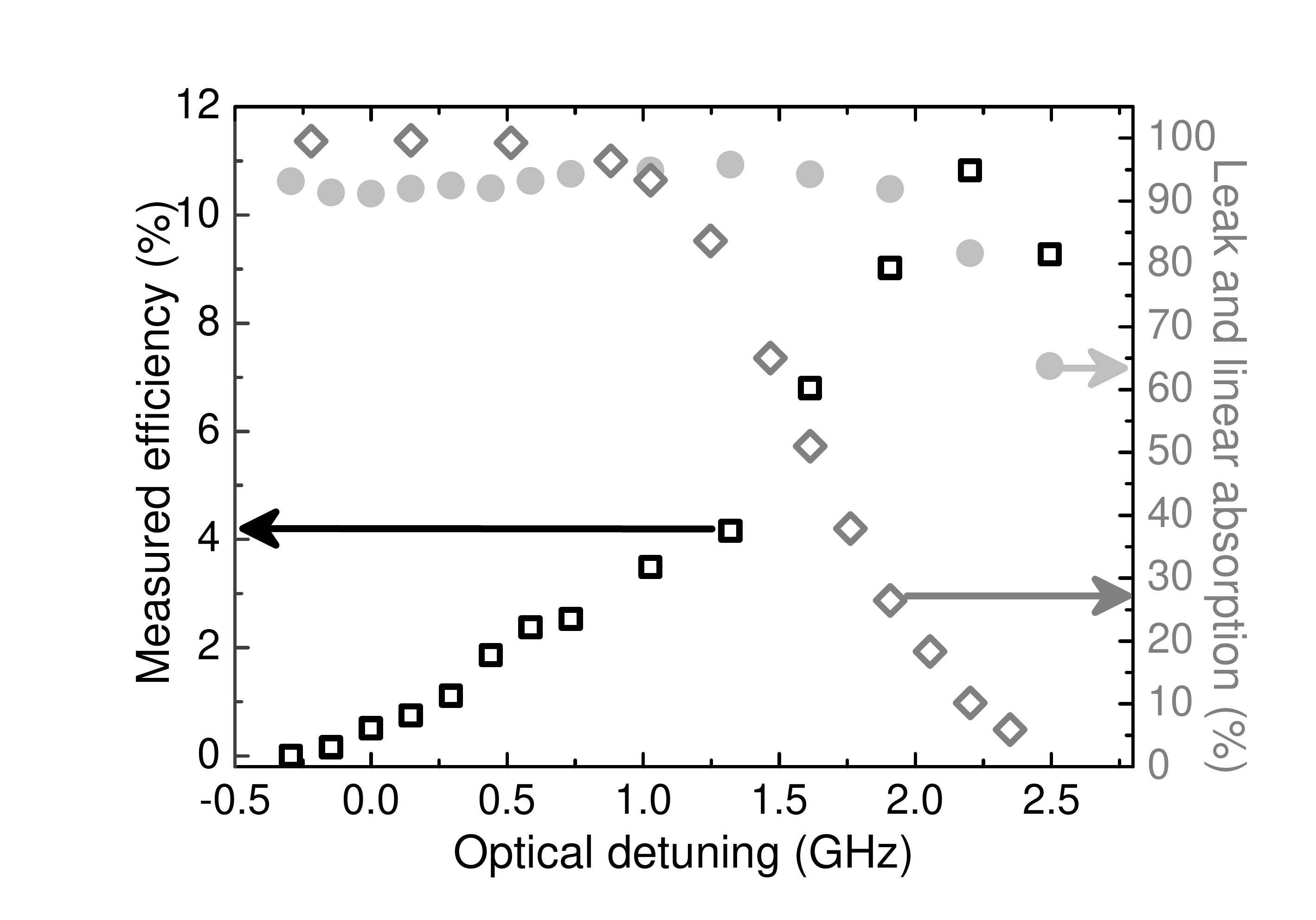} 
\caption{\label{fig2} Storage efficiency (open squares), 1-the leak level (dots) and linear absorption of a cw probe in the absence of the coupling beam (open diamonds) as a function of the optical detuning of the coupling beam for a 2\,$\mu$s rise time exponential probe pulse. The leak level is defined as the leak area divided by the incoming pulse area. A positive detuning corresponds to a laser frequency larger than the transition frequency. The storage time is 3\,$\mu$s.}
\end{figure}

\begin{figure}
\center
\includegraphics[width=0.4\textwidth]{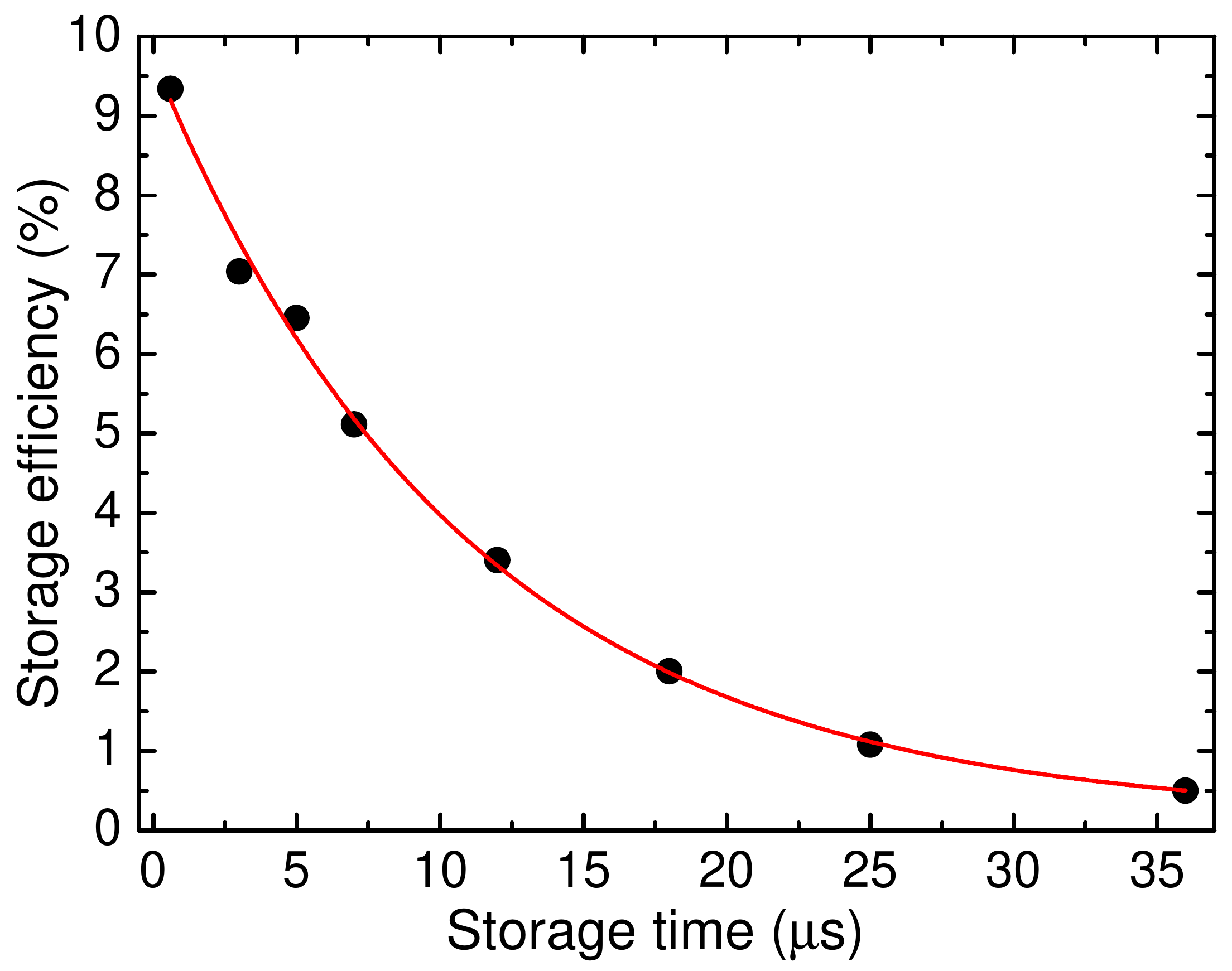} 
\caption{\label{fig3}Measured evolution of the storage efficiency as a function of the storage time $\tau$ for an optical detuning of about 1.5\,GHz and a 4\,$\mu$s rise-time exponential probe pulse. The continuous line is an exponential fit with a decay time of $11\ \mu\mathrm{s}$.}
\end{figure}

Storage efficiency results are reported in figs.\,\ref{fig2} and \ref{fig3}. Figure \ref{fig2} shows the evolutions of the storage efficiency versus optical detuning of the coupling and probe beams with respect to the center of the Doppler profile. The weak probe pulse  has a 2\,$\mu$s rise time exponential edge, and the storage time is fixed and equals 3\,$\mu$s. A positive detuning corresponds to a laser frequency larger than the transition frequency: on this side of the transition, the next level ($^3$P$_0$) is nearly 30\,GHz away. On the other side, i.\,e.\,for negative detunings, the $^3$P$_2$ level is only 2.29\,GHz away. This is why we focus here mainly on positive detunings. The maximum efficiency -- about 11\% for a 3\,$\mu$s storage time -- is obtained for a detuning of 2.2\,GHz. For such a detuning, the efficiency at zero storage time was measured to be more than 15\%. It is not as high as the ones obtained in other systems e.g. rubidium at room temperature \cite{Novikova2012} or cold atoms \cite{Chen2013}, but we would like to stress the fact that the absorption depth $\alpha L$ of the medium is less than 7and thus the optical depth $d=\alpha L/2$ is less than 3.5 in the notations used by Gorshkov \& al \cite{Gorshkov2007b}. Such an optical depth gives less than 30\% maximum theoretical efficiency in the forward direction according to ref. \cite{Gorshkov2007b}.. Knowing that the pumping efficiency decreases with optical detuning and that our maximum coupling power was limited to 20\,mW at the entrance of the cell, the efficiencies that we reach here are not far from the theoretical maximum.\\
One might wonder if the origin of a maximum transmission obtained for a detuning of twice to thrice the Doppler width can be found in the breakdown of the optimal adiabatic storage. Gorshkov \& al \cite{Gorshkov2007b} have shown that in the adiabatic limit, any input mode can be stored with the same maximum efficiency, independently of the optical detuning. The necessary and sufficient condition on the duration $T$ of the probe pulse to verify the adiabatic approximation is $Td\gamma\gg1$, where $\gamma$ is the polarization decay rate. It is when the probe adiabaticity is not valid that the efficiency increases with the optical detuning. Nevertheless, in our conditions, $\gamma\approx1.4\times10^8$\,s$^{-1}$, which gives several hundreds for $Td\gamma$: for such values, the theoretical plots given by Gorshkov \& al \cite{Gorshkov2007b} show that the maximum efficiency is reached whatever the optical detuning is.

Figure \ref{fig2} also shows the leak level (to be more precise 1- the leak level) and the absorption of a linear cw probe with respect to the optical detuning. The leak level is obtained by the area of the leak pulse divided by the area of the incoming pulse. It remains nearly constant for small detunings while the efficiency quickly increases. The linear absorption decreases for optical detunings larger than 1\,GHz, which is consistent with a 0.9\,GHz Doppler broadening. The fact that the leak level remains flat when the optical detuning increases is consistent with the increase of the efficiency if it is linked to a better mapping of the light on the spin wave. A full theoretical treatment is still to be done and should take into account the absorption of the coupling beam along the cell.

Figure \ref{fig3} shows the exponential decrease of the storage efficiency with the storage time. These data are obtained for an optical detuning of about 1.5\,GHz with weak probe pulses having an exponential leading edge with a 4\,$\mu$s rise time. The decay time of the exponential evolution of fig.\,\ref{fig3} is 11 $\mu$s, which corresponds to $\Gamma_R/2\pi\simeq 14$\,kHz, where $\Gamma_R$ is the Raman coherence decay rate. This value is consistent with measurements of the evolution of the width of the EIT transparency window with respect to the coupling power \cite{Goldfarb2008}, that give $\Gamma_R/2\pi\simeq14$\,kHz. This is broader than the few kHz that we had previously recorded \cite{Ghosh2009}, as expected from the fact that we now have reduced beam diameters down to 3\,mm, thus decreasing the transit time.

\begin{figure}
\center
\includegraphics[width=0.45\textwidth]{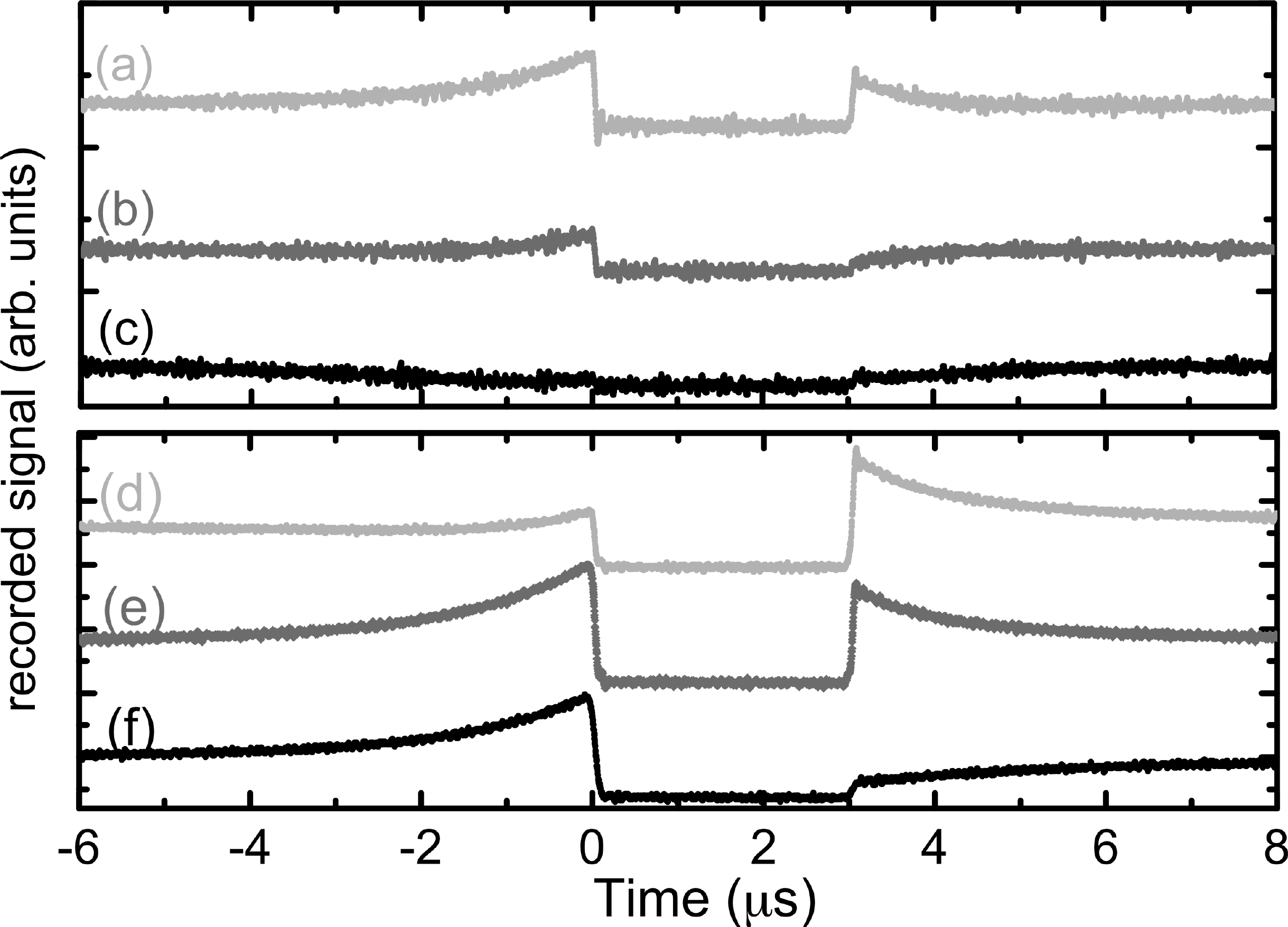} 
\caption{\label{fig4} Traces recorded at different times when the laser frequency is (a-c) close to resonance and (d-f) at a frequency 2.2\,GHz larger than the center of the Doppler profile. The retrieved pulse is in phase with the leak in the case (a-c) while is is clearly phase shifted in the case (d-f).}
\end{figure}

The homodyne detection that we use makes the recorded leak and signal amplitudes oscillate when the phase of the coupling beam is scanned. This permits us to measure the phase of the retrieved pulse with respect to the leak of the incident pulse, for different values of the detuning. Figures \ref{fig4}(a,b,c) reproduce the recorded temporal traces for three values of the phase of the local oscillator, when the frequency of the laser is close to the center of the Doppler profile. These traces clearly show that the leak and the retrieved pulse interfere in the same way with the local oscillator. For example, in fig.\,\ref{fig4}(a), the interference is constructive for both pulses, while in fig.\,\ref{fig4}(c), the interference is destructive for both pulses. This shows that, in the case of figs.\,\ref{fig4}(a-c), the retrieved pulse is in phase with the incident pulse. In contrast, in figs.\,\ref{fig4}(d,e,f) showing similar traces recorded for a 2.2\,GHz optical detuning, a phase shift between the leak and the retrieved pulse is clearly visible, as the interferences can, for example, be constructive for the leak signal and destructive for the retrieved pulse [see in particular fig.\,\ref{fig4}(f)]. It must be emphasized here that we do not see any dependence of this phase shift on the storage time. Moreover, it is worth mentioning that the scan of the piezoelectric actuator is performed slowly enough to ensure that the relative phase between the coupling and probe beams remains constant over the duration of the incoming probe pulse and during the reemission of the stored pulse. Indeed, the scanning frequency was 20\,mHz for an amplitude of the order of 5 times $2\pi$.

\begin{figure}
\center
\includegraphics[width=0.45\textwidth]{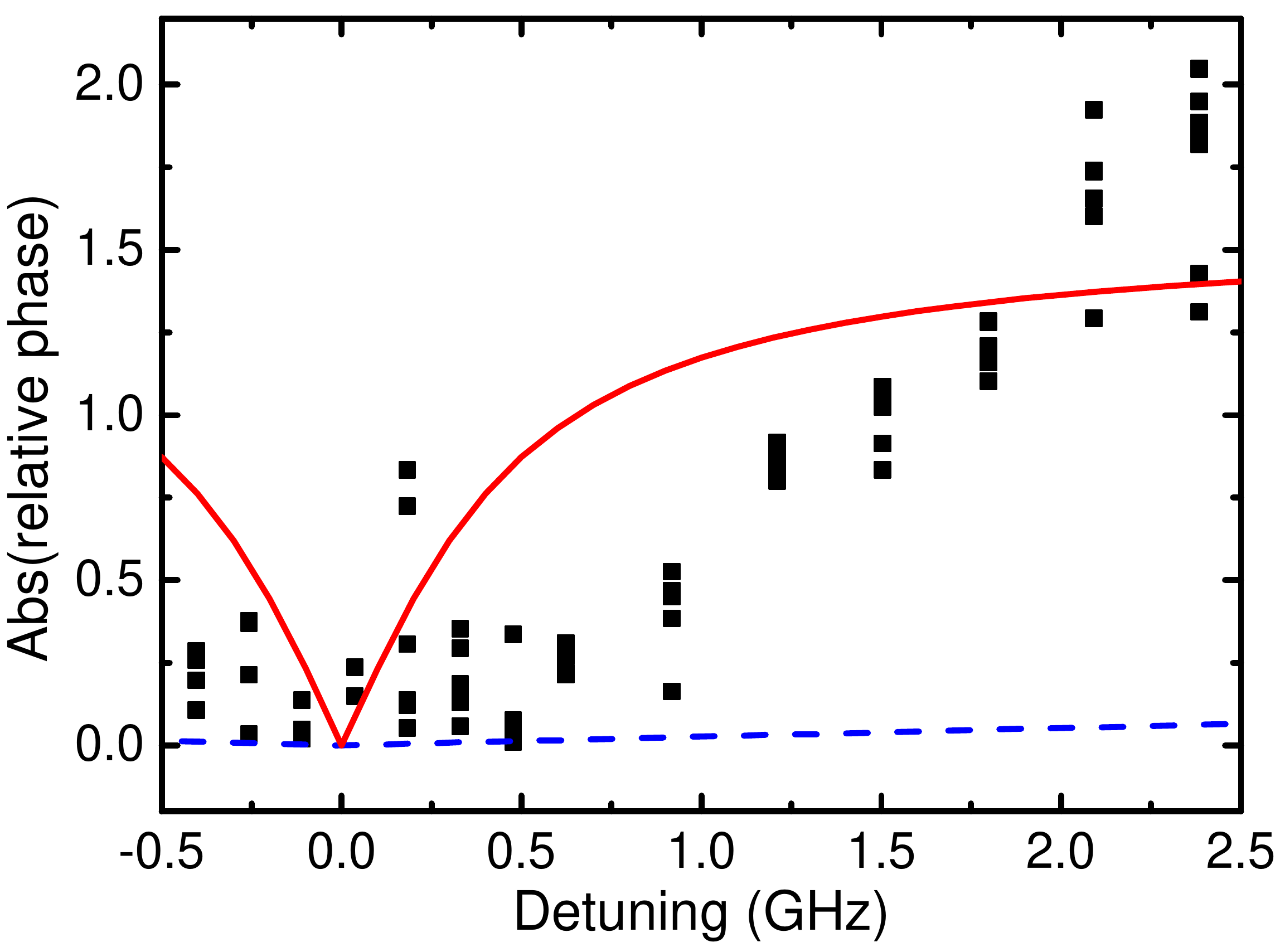} 
\caption{\label{fig5}Evolution of the magnitude of the phase shift acquired during storage and retrieval versus optical detuning. The squares are experimental measurements and the continuous and dashed lines are obtained by extracting $|\varphi_{EIT}|$ from eq. (\ref{eq02}): the continuous red line is plotted with $\Gamma/2\pi=0.4$\,GHz and a coupling power down to zero, while the dashed blue line is plotted with $\Gamma/2\pi=0.42$\,GHz, $\Gamma_R/2\pi=14$\,kHz and $\Omega_C/2\pi=23$\,MHz.}
\end{figure}
We could not notice any dependence of this phenomenon on the storage time, and it is thus clear from the observations in fig. \ref{fig4} that the storage process itself introduces an extra phase shift to the retrieved light pulse, when it is  performed with a non-zero detuning. The phase shift reported here is completely different from the ones previously observed for Raman detuned EIT storage and a coupling beam at optical resonance \cite{Chen2005a,Chen2006} and from the control of the stored phase using a pulsed magnetic field \cite{Mair2002} .

In order to understand the origin of this phase shift acquired during storage, we attempt a perturbative treatment of the system. We consider the usual equation of evolution of the density matrix of the three-level system, neglecting all population terms except the one in the ground state sublevel $m=1$ ($\sigma_{1,1}\simeq1$). If $\Omega_C\gg\Omega_P$, we obtain the following expression for the steady-state Raman coherence between the Zeeman ground levels $m=\pm 1$ in the presence of both the coupling and probe beams:
\begin{equation}
  \tilde{\sigma}_{1,-1}=\frac{-\Omega_C\Omega_P^*}{(\Gamma_R+i\delta_R)(\Gamma+i\Delta_P)+|\Omega_C|^2}\ ,\label{eq01}
\end{equation}
where $\Gamma$ is the optical coherence decay rate, $\Delta_P$ the probe optical detuning and $\delta_R$ the Raman detuning. $\tilde{\sigma}_{1,-1}$ is expressed in the rotating frame, with $\tilde{\sigma}_{1,-1}=\sigma_{1,-1}\exp{(i\delta_R t)}$. At Raman resonance, one has $\tilde{\sigma}_{1,-1}=\sigma_{1,-1}$.

At Raman resonance, this equation shows that the pulse stored in the Raman coherence has acquired an extra phase shift. This phase shift is given by the argument $\varphi_{EIT}$ of the Raman coherence of eq. (\ref{eq01}), given by: 
\begin{equation}
\tan\varphi_{EIT}=\frac{\Gamma_R\Delta_P}{|\Omega_C|^2+\Gamma_R \Gamma}\ .\label{eq02}
\end{equation}
with $\Delta_P=\Delta_C$ at Raman resonance.

The second phase of the protocol is the retrieval process, for which the Hamiltonian to be considered reads, in the rotating frame, 
\begin{equation}
\hat{H}=\hbar\left(\begin{array}{ccc}
0 & \Omega_{C2} & 0 \\          
\Omega_{C2}^* & 0 & 0 \\ 
0 & 0 & 0 \\           
\end{array}\right)\ .
\end{equation}
This Hamiltonian is written here in the basis $\{|e\rangle,|-1\rangle,|1\rangle\}$, where $|e\rangle$ is the $m=0$ sublevel of the excited level, and $|\pm1\rangle$ are the $m=\pm1$ sublevels of the ground state. We can then perform a first-order perturbative calculation of the retrieved field amplitude, with an initial density matrix, i.\,e., just before the coupling field is switched on again, given by:
\begin{equation}
\sigma_{ini}=\left(\begin{array}{ccc}
0 & 0 & 0 \\          
0& 0 & \sigma_{-1,1} \\ 
0 & \sigma_{1,-1} & 1 \\           
\end{array}\right)\ .
\end{equation}
It should be noted that the coupling Rabi frequency does not have to be the same for the storage and retrieval processes, neither its modulus, nor its phase. Using the von Neumann equation, $i\hbar d\sigma/dt=[\hat{H},\sigma]$ with the initial condition $\sigma=\sigma_{ini}$, we find that $d\sigma_{1,e}/dt=i\Omega_{C2}^*\sigma_{1,-1}$. As the retrieved signal amplitude $\Omega_{r}$ is proportional to $i\sigma_{1,e}^*$, its phase is found to be $\varphi_{C2}-\arg{\sigma _{1,-1}}$, where $\varphi_{C2}$ is the phase of the coupling field used for the restitution. As we use the reading beam $\Omega_{C2}$ as a local oscillator for the homodyne detection, the phase $\varphi_{C2}$ is eliminated, and we obtain a phase shift $\varphi_{EIT}$ for the retrieved signal compared to the leak signal. Furthermore, we notice that the same result could be obtained even with a readout coupling beam at a frequency different from the storage coupling beam. In all cases, the retrieved pulse will be emitted at the frequency of the readout coupling beam with the stored phase shift $\varphi_{EIT}$.

Figure \ref{fig5} shows the magnitude of the measured phase shift as a function of the optical detuning. The theoretical continuous and dashed lines correspond to $\left|\arctan(\Gamma_R\Delta_P/(|\Omega_C|^2+\Gamma_R \Gamma))\right|$, as expected from eq. (\ref{eq02}). Following Gorshkov \& al \cite{Gorshkov2007c}, we replaced the homogeneous line width by the Doppler width and used $\Gamma/2\pi=0.4$\,GHz given by the effective Doppler width and $\Gamma_R/2\pi=14$\,kHz.  The dashed blue line is plotted for $\Omega_C/2\pi=23$\,MHz corresponding to the average of the coupling power along the cell, while the continuous red line is obtained for zero coupling power. The agreement between theory and experiment seems good at zero coupling power, but not when we take into account the existence of the coupling beam that was indeed used for storage. A proper theoretical model remains to be developped, to take into account propagation and transient effects that might explain the fact that a simple steady-state first order model fails to reproduce the data.

In conclusion, we have shown that room-temperature metastable $^4$He permits one to achieve light storage with an efficiency larger than 10\% for a storage time of 3\,$\mu$s, and with a decay time of the storage efficiency equal to $11\ \mu \mathrm{s}$. The efficiency is shown to be maximum for detunings of the order of twice the Doppler linewidth. Moreover, we have observed the existence of an extra phase shift accumulated by the retrieved pulse during the storage process. This phase shift is explained by the relative phase shift recorded by the Raman coherence, which is created by the incoming probe pulse in the presence of the coupling beam. The theoretical model based on a steady-state first-order-of-perturbation treatment of the evolution of the density matrix of the system does not give satisfying results and a full theoretical treatment that takes into account transient phenomena is still to be done. These results are interesting as some of the best light storage efficiencies were obtained with a gradient echo protocol in similarly detuned regimes for gas cells at room temperature \cite{Hosseini2009,Hosseini2011}, and single photon level low noise storage experiments have been performed with good efficiencies in large detunings conditions in caesium cells \cite{Reim2011}. Moreover, such an optical control of the phase of the retrieved pulse could be used to design phase gates for quantum computing applications.

\acknowledgments
This work was performed in the framework of the CNRS (France)-DST(India) ``Programme International de Coop\'eration Scientifique". The work of S. K. was supported by the Council of Scientific and Industrial Research, India. The work of M.-A. M. was supported by the Delegation Generale de l'Armement (DGA), France.


\end{document}